\documentclass[twocolumn,aps,superscriptaddress,showpacs]{revtex4}
\usepackage{amssymb}
\usepackage{amsmath}
\usepackage{mathrsfs}
\usepackage{graphicx}
\usepackage{subfigure}
\usepackage[normalem]{ulem}
\usepackage[dvips]{color}
\usepackage{academicons} 
\usepackage{xcolor}
\usepackage{hyperref}

\hypersetup{hypertex=true,
colorlinks=true,
linkcolor=magenta,
citecolor=magenta,
urlcolor=cyan} 
\definecolor{orcidgreen}{HTML}{A6CE39}
\newcommand{\orcid}[1]{\href{https://orcid.org/#1}{\textcolor{orcidgreen}{\aiOrcid}}}

\begin{document}
\title{Temperature fluctuations in a realistic Polyakov-loop extended Nambu--Jona-Lasinio Model along the freeze-out line}
\author{He Liu \orcid{0000-0002-8658-2851}}
\email{liuhe@qut.edu.cn}
\affiliation{Science School, Qingdao University of Technology, Qingdao 266520, China}
\affiliation{The Research Center of Theoretical Physics, Qingdao University of Technology, Qingdao 266033, China}
\author{Peng Wu}
\affiliation{Science School, Qingdao University of Technology, Qingdao 266520, China}
\affiliation{The Research Center of Theoretical Physics, Qingdao University of Technology, Qingdao 266033, China}
\author{Hong-Ming Liu \orcid{0000-0002-6479-9785}}
\email{liuhongming13@126.com}
\affiliation{Science School, Qingdao University of Technology, Qingdao 266520, China}
\affiliation{The Research Center of Theoretical Physics, Qingdao University of Technology, Qingdao 266033, China}
\author{Peng-Cheng Chu \orcid{0000-0001-7311-9684}}
\email{kyois@126.com}
\affiliation{Science School, Qingdao University of Technology, Qingdao 266520, China}
\affiliation{The Research Center of Theoretical Physics, Qingdao University of Technology, Qingdao 266033, China}
\date{\today}

\begin{abstract}
We extend and reparameterize the Polyakov--Nambu--Jona-Lasinio (PNJL) model to reproduce lattice simulation data at zero baryon chemical potential and to position its critical endpoint (CEP) within the BES-II experimental energy range. Using this realistic PNJL model, we investigate the behavior of the second-order temperature cumulant $C_2$ along the freeze-out line, aiming to understand the non-monotonic energy dependence of the two-particle transverse momentum correlation $C_{p_T}$ observed by the STAR Collaboration. Our results show a distinct dip structure in $C_2$ near the CEP and the first-order phase boundary on the phase diagram. Along the experimentally extracted freeze-out line, the dip minimum occurs around 7.7 GeV, and its trend is consistent with that observed by STAR, suggesting that the non-monotonic dependence of $C_{p_T}$ may be related to the CEP. Our results also indicate that cumulant ratios such as $C_3/C_2^2$ or $C_4/C_2^3$ eliminate the influence of initial volume fluctuations and may better reveal the underlying critical fluctuations. Further verification could be pursued through hydrodynamic or transport simulations that incorporate critical dynamics. These results and predictions may provide an a priori theoretical basis for future experimental measurements of higher-order event-mean transverse momentum fluctuations.
\end{abstract}

\maketitle

\textit{Introduction}. Investigating the phase structure of Quantum Chromodynamics (QCD) is a primary goal of relativistic heavy-ion collision experiments. To search for the critical endpoint (CEP) and the first-order phase transition boundary, the Beam Energy Scan (BES) program at the Relativistic Heavy Ion Collider (RHIC) systematically probes regions of baryon chemical potential ($\mu_B$) up to approximately 760 MeV by varying the collision energy~\cite{Bzdak_mapping_2020,Chen_properties_2024,Abdallah_flow_2021}. Using high-statistics data from the BES-II program, the STAR Collaboration has recently performed precise measurements of observables sensitive to the CEP and first-order phase transition. For example, the cumulant ratio of net-proton number, $C_4/C_2$, exhibits a minimum near $\sqrt{s_{NN}} = 19.6$ GeV in 0-5\% central collisions relative to a non-critical baseline, with a statistical significance of $2-5\sigma$~\cite{Aboona_precision_2025}. In recent fixed-target Au+Au collisions, the STAR Collaboration reported the first systematic measurements of two-particle transverse momentum correlations $C_{p_T} = \sqrt{\langle \Delta p_{T,i} \Delta p_{T,j} \rangle}/{\langle \langle p_T \rangle \rangle}$ in the high-baryon-density region ($\mu_B\approx 760-400$ MeV) at center-of-mass energies $\sqrt{s_{NN}} = 3.0-7.7$ GeV from BES-II data, revealing a significant non-monotonic dip structure in the energy dependence for central collisions~\cite{STAR_non-monotonicity_2026}. These observed non-monotonic variations may be sensitive to the presence of the QCD CEP~\cite{Aboona_precision_2025,STAR_non-monotonicity_2026}.

Hydrodynamic simulations reveal a simple proportionality between the average transverse momentum $\langle p_T \rangle$ of final-state particles and the effective temperature $T_{\text{eff}}$~\cite{Gardim_thermodynamics_2020}. Using the relation $c_s^2 = dp/d\varepsilon=\frac{d\ln T}{d\ln s}|_{T_{\text{eff}}} \thicksim d\ln \langle p_T \rangle / d\ln N_{ch}$, where $c_s$ is the speed of sound and $N_{ch}$ is the charged-particle multiplicity, the CMS Collaboration measured $T_{\text{eff}} = 219\pm8$ MeV and $c_s^2 = 0.241\pm0.016$ from Pb+Pb collisions at \(\sqrt{s_{NN}}=5.02\) TeV~\cite{Hay_extracting_2024}. In high-energy nuclear-nuclear collisions, the sensitivity of the fluctuations of the event by event mean $\langle p_T\rangle$ (i.e., $\langle(\delta p_T)^2\rangle$ is also found to be sensitive to nuclear deformation for imaging of nuclear shapes~\cite{Abdulhamid_imaging_2024,STAR_imaging_2025}. Consequently, transverse momentum fluctuations can be used to extract information about the dynamical evolution of the collision system. A recent study introduced a novel thermodynamic state function that establishes an analytical connection between temperature and $\langle p_T \rangle$ fluctuations, providing analytical expressions for cumulants of temperature fluctuations at arbitrary orders~\cite{Chen_high-order_2025}. In our previous work based on the Polyakov-loop extended Nambu–Jona-Lasinio (PNJL) model~\cite{Liu_fluctuations_2026}, we further demonstrated that the higher-order cumulant ratios of temperature fluctuations exhibit distinctly different non-monotonic features across various phase transition regions, including the chiral crossover, the first-order phase transition, the vicinity of the CEP, and the deconfinement phase transition.

Although the STAR experiment reported that the charged-particle transverse momentum correlation $C_{p_T}$ is non-monotonic with a pronounced minimum around $\sqrt{s_{NN}} \approx 7.7$ GeV, a recent study suggested that this minimum can be understood as an effect of mixing two different particle distributions (especially protons and pions) during the transition from a baryon-dominated to a meson-dominated system in heavy-ion collisions, and questioned the interpretation of the observed minimum as a signal of the CEP~\cite{Reichert_non-monotonicity_2026}. Given the linear relationship between $T$ and $\langle p_T \rangle$, the authors of Ref.~\cite{Chen_high-order_2025} calculated temperature fluctuations along the chemical freeze-out line using the 2+1 flavor LEFT-fRG framework, but did not observe a non-monotonic dip structure in the second-order cumulant of temperature $C_2$ as a function of collision energy. To investigate whether the non-monotonic behavior of $\langle p_T \rangle$ fluctuations is intrinsically related to the QCD CEP, we employ an extended PNJL model to systematically study the temperature fluctuations along the freeze-out line determined by the BES experiment, as well as their connection to the QCD phase structure, aiming to provide testable theoretical predictions for future heavy-ion collision experiments.

\newcommand{\thickhline}{
\noalign{\hrule height 2\arrayrulewidth}
}
\begin{table}
\centering 
\caption{Parameters for the Polyakov-loop potential in the realistic PNJL model.}
\begin{tabular*}{0.48\textwidth}{@{\extracolsep{\fill}} *{7}{c}} 
\thickhline
$T_0$(MeV) & $a_0$ & $a_1$ & $a_2$ & $b_3$ & $b_4$ & $\kappa$ \\ 
\hline 
175 & 6.75 & -8.1 & 0.26 & 0.805 & 7.555 & 0.02\\ 
\thickhline
\end{tabular*}
\end{table}

\textit{Theoretical model}. Theoretical investigations of the QCD phase structure have been performed using first-principles approaches (e.g., Lattice QCD simulations~\cite{Aoki_the_2006,Gupta_scale_2011,Bazavov_chiral_2012,Bazavov_equation_2014,Borsanyi_full_2014,Bellwied_the_2015,Bazavov_chiral_2019,Ding_chiral_2019} and functional methods~\cite{Braun_phase_2011,Gao_chiral_2021,Fischer_qcd_2019,Fu_qcd_2020,Braun_chiral_2020}) and low-energy effective models~\cite{Schwarz_phase_1999,Fukushima_phase_2008,Liu_isospin_2016}. Lattice simulations indicate that near zero baryon chemical potential ($\mu_B$=0), the phase transition between the quark-gluon plasma (QGP) and the hadron resonance gas (HRG) is a smooth crossover~\cite{Aoki_the_2006,Gupta_scale_2011}. The critical temperature $T_c$ for the crossover transition is in the range of 150--160 MeV, as reported by the HotQCD~\cite{Bazavov_chiral_2012,Bazavov_equation_2014,Bazavov_chiral_2019} and Wuppertal-Budapest (WB)~\cite{Borsanyi_full_2014,Bellwied_the_2015} collaborations. Functional methods such as the Dyson–Schwinger equation (DSE) and the functional renormalization group (fRG)~\cite{Fu_qcd_2020,Gao_chiral_2021} yield consistent results with $T_c \approx 157$ MeV. At finite/large $\mu_B$, lattice results are no longer available. Studies using advanced functional methods or effective models such as the NJL-type models suggest that this transition may become first-order, connected to the crossover region via a critical endpoint~\cite{Fukushima_phase_2008,Liu_isospin_2016,Fu_qcd_2020,Gao_chiral_2021}. However, the predicted location of the CEP differs between fRG and DSE approaches~\cite{Fu_qcd_2020,Gao_chiral_2021}. Effective models such as the NJL/PNJL model, despite possessing similar global symmetries to QCD (e.g., chiral symmetry), exhibit large variations in both $T_c$ and the CEP position. 

\begin{figure}[tbh]
\includegraphics[scale=0.39]{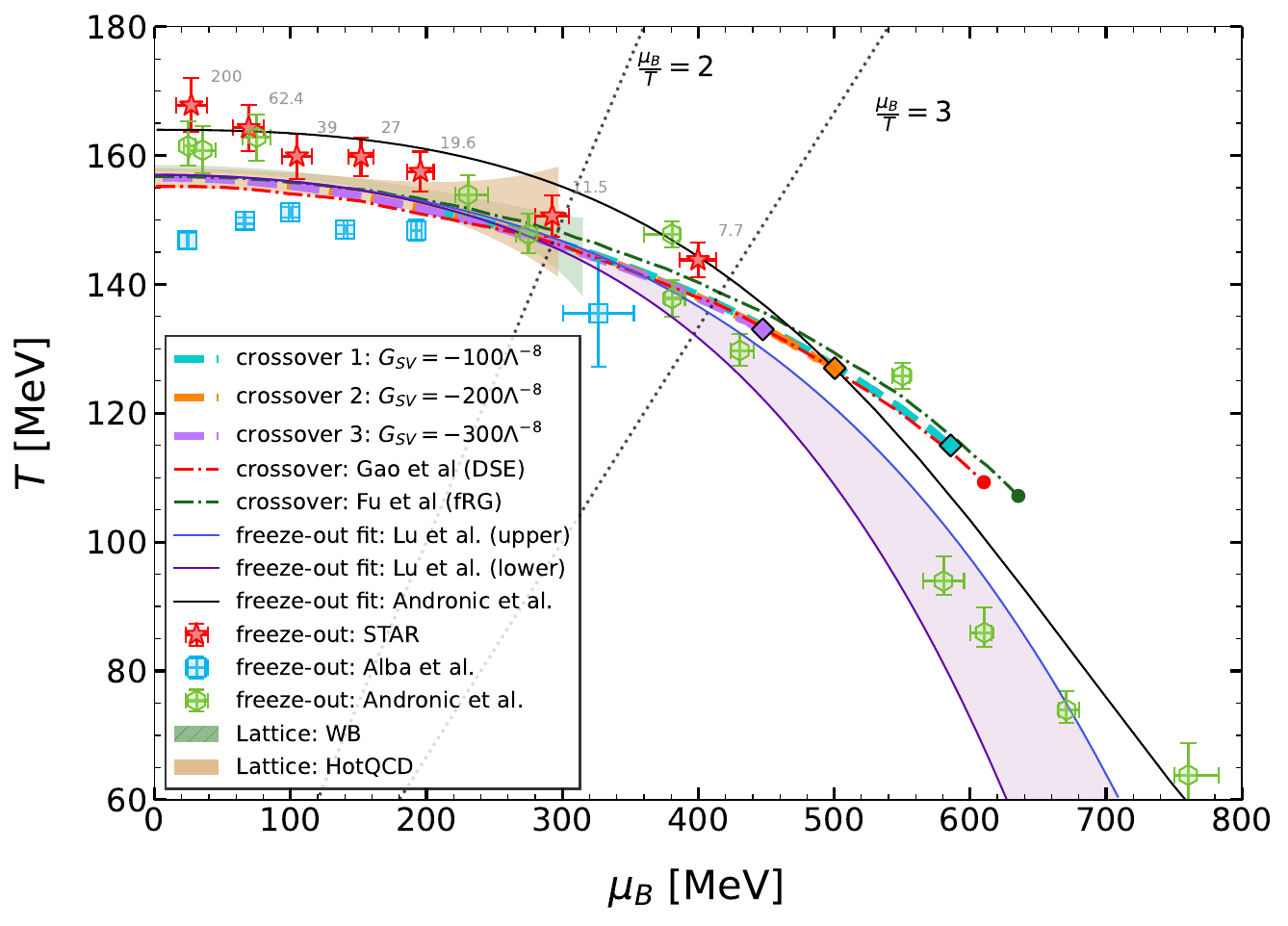}
\caption{QCD phase diagram in the $\mu_B-T$ plane for different $G_{SV}$ based on the realistic PNJL model. The dashed lines denote the chiral crossover, and the solid diamonds mark the CEPs. Results from functional methods (fRG~\cite{Fu_qcd_2020} and DSE~\cite{Gao_chiral_2021}) and lattice simulations (WB~\cite{Bellwied_the_2015} and HotQCD~\cite{Bazavov_chiral_2019}) are shown for comparison. We also include the experimental chemical freeze-out data (STAR~\cite{Adamczyk_bulk_2017}, Alba et al.~\cite{Alba_freeze-out_2014}, Andronic et al.~\cite{Abdallah_flow_2021}) and freeze-out fitting lines (Lu et al.~\cite{Lu_extracting_2026}, Andronic et al.~\cite{Andronic_the_2010}).} \label{fig1}
\end{figure}

\begin{figure*}[htbp]
    \centering 
    \begin{minipage}[t]{0.45\textwidth}
        \centering
        \includegraphics[trim=20 20 80 50, clip, width=\linewidth]{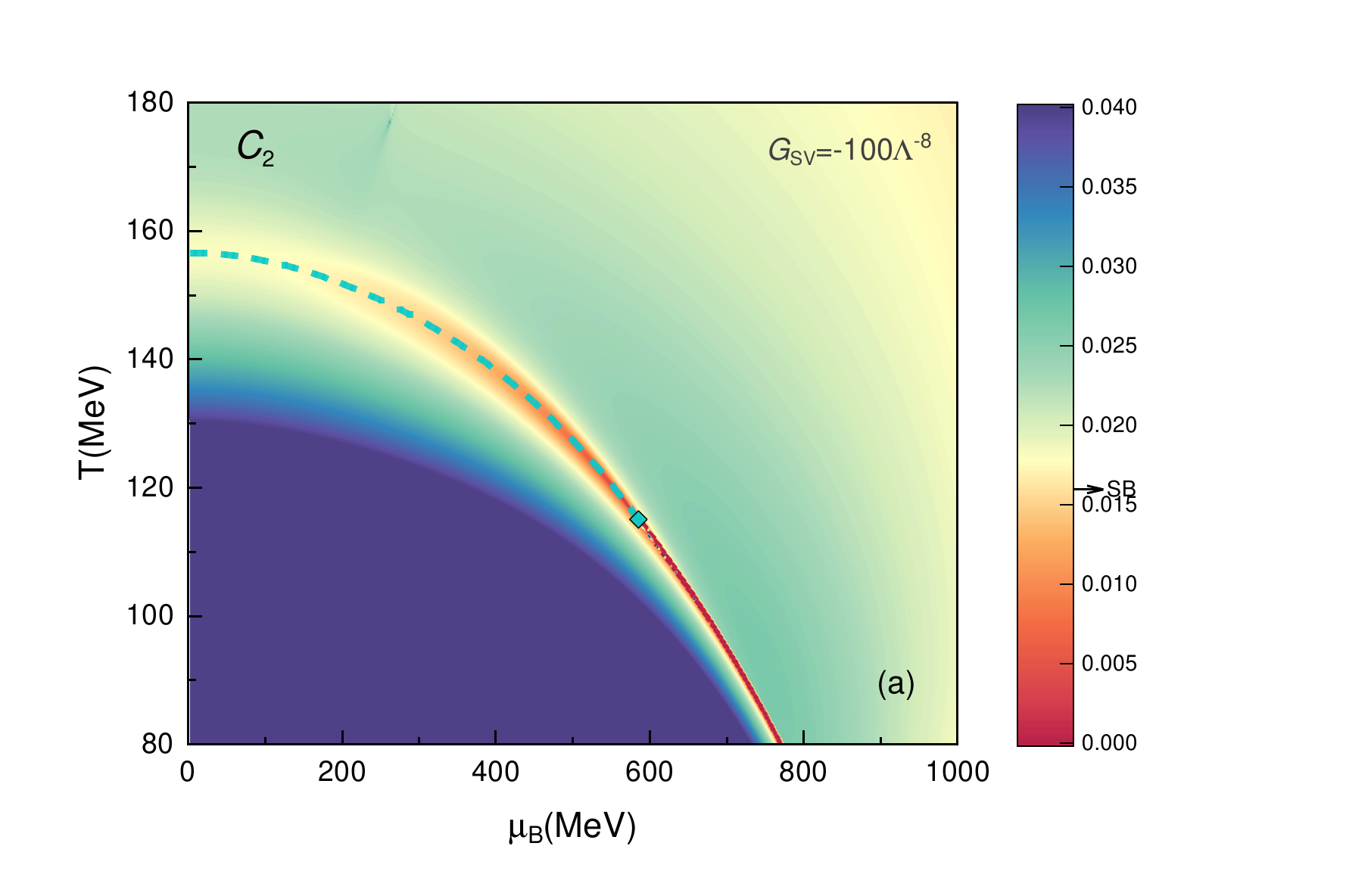}
    \end{minipage}
    \begin{minipage}[t]{0.405\textwidth}
        \centering
        \includegraphics[width=\linewidth]{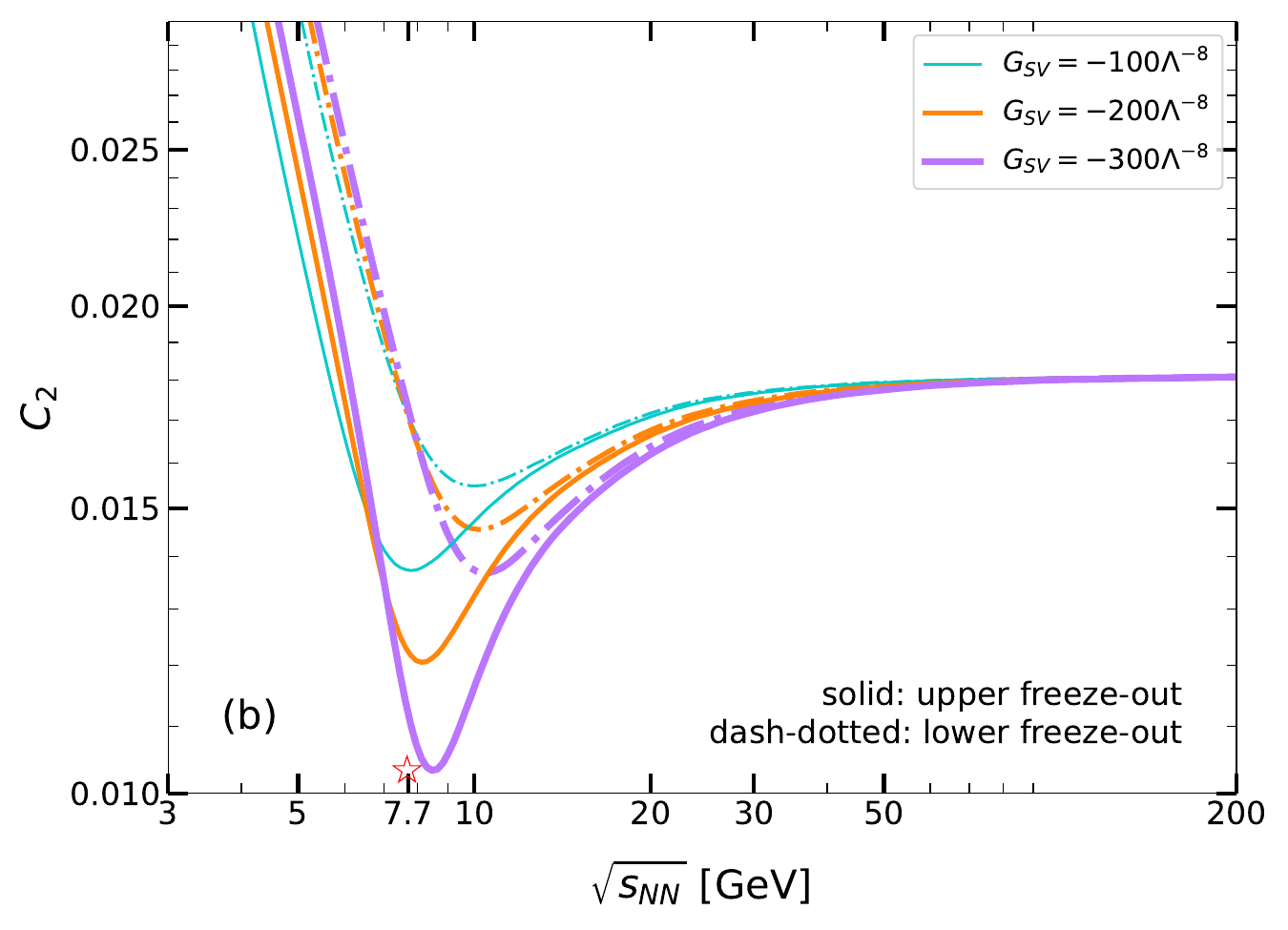}
    \end{minipage}
    \caption{Panel (a): the contour map of temperature cumulant $C_2$ with $G_{SV}=-100\Lambda^{-8}$ in the full $\mu_B$-$T$ phase diagram. The dashed curve denotes the chiral crossover, and the diamond marks the critical endpoint. Panel (b): $C_2$ along the experimental chemical freeze-out lines for various $G_{SV}$. The solid and dash-dotted lines represent the upper and lower freeze-out lines, respectively, as extracted from Ref.~\cite{Lu_extracting_2026}.}\label{fig2}
\end{figure*}

In view of this, we extend and reparametrize the PNJL model by introducing two eight-quark interaction terms to reproduce the lattice QCD data at zero baryon chemical potential, and by incorporating a vector-scalar coupling interaction to adjust the CEP position to accommodate experimental results. The Lagrangian density of this realistic 3-flavor PNJL model is given by~\cite{Bhattacharyya_reparametrizing_2017,Bhattacharyya_finite_2020,Li_the_2019,Sun_qcd_2021,Yang_properties_2024}
\begin{eqnarray}\label{eq1}
\mathcal{L}_{\textrm{PNJL}}^{\textrm{SU(3)}} &=& \bar{\psi}(i\gamma^{\mu}D_{\mu}\!-\!\hat{m})\psi 
\!+\!G_{S}\sum_{a=0}^{8}[(\bar{\psi}\lambda_{a}\psi)^{2}\!+\!(\bar{\psi}i\gamma_{5}\lambda_{a}\psi)^{2}] 
\notag\\
&-& K\{\det[\bar{\psi}(1+\gamma_{5})\psi]+\det[\bar{\psi}(1-\gamma_{5})\psi]\}
\notag\\
&+&\mathcal{L}_1^{8q}+\mathcal{L}_2^{8q}+\mathcal{L}_{SV}+\mathcal{U} ^{\prime}(\Phi[A],\bar{\Phi}[A],T),
\end{eqnarray}
where $\psi = (u, d, s)^T$ represents the 3-flavor quark field, $\hat{m} = \mathrm{diag}(m_u, m_d, m_s)$ is the current quark mass matrix, and $\lambda_a$ are the Gell-Mann matrices of flavor SU(3), with $\lambda_0 = \sqrt{2/3}I$. The covariant derivative is defined as $D_{\mu}=\partial_{\mu}-iA_{\mu}$, where the background gluon field $A_{\mu}=\delta_{\mu,0}A_0$ is taken to be constant and uniform. The temperature-dependent Polyakov effective potential $\mathcal{U}'(\Phi[A],\bar{\Phi}[A],T)$ is a function of the Polyakov loop $\Phi[A]$ and its Hermitian conjugate $\bar{\Phi}[A]$, and its explicit form is given in Refs.~\cite{Bhattacharyya_reparametrizing_2017}:
\begin{eqnarray}\label{eq2} 
\frac{\mathcal{U}^{\prime}(\Phi,\bar{\Phi},T)}{T^4}=\frac{\mathcal{U}(\Phi,\bar{\Phi},T)}{T^4}-\kappa \ln[J(\Phi,\bar{\Phi})].
\end{eqnarray}
Here $\mathcal{U}(\Phi,\bar{\Phi},T)$ is chosen as a Landau-Ginzburg-type potential compatible with the global $Z(3)$ symmetry of the Polyakov loop; the term $J(\Phi,\bar{\Phi})=1-6\Phi\bar{\Phi}+4(\Phi^3+\bar{\Phi}^3)-3(\Phi\bar{\Phi})^2$ is the Jacobian of transformation from the Polyakov loop to its traces~\cite{Bhattacharyya_finite_2020}, and $\kappa$ is a dimensionless parameter. The effective potential $\mathcal{U}(\Phi,\bar{\Phi},T)$ is given by
\begin{eqnarray}\label{eq3} 
\frac{\mathcal{U}(\Phi,\bar{\Phi},T)}{T^4}=-\frac{b_2(T)}{2}\Phi\bar{\Phi}-\frac{b_3}{6}(\Phi^3+\bar{\Phi}^3)+\frac{b_4}{4}(\Phi\bar{\Phi})^2,
\end{eqnarray}
where the temperature-dependent coefficient $b_2(T)$ is $b_2(T)=a_0+a_1 \frac{T_0}{T} \textrm{exp} (-a_2 \frac{T}{T_0})$; $b_3$ and $b_4$ are chosen to be constants. In the high-temperature limit, the Polyakov loop field is expected to approach unity. Therefore, for the effective model of pure gluon theory, the minimization condition requires that $\Phi = 1$ as $T \to \infty$, and that the pressure reduces to that of a massless gluon gas. Using these two conditions, the parameters $b_3$ and $b_4$ can be determined by $b_2(T\to\infty) = a_0$. Subsequently, the parameters $a_1, a_2, T_0$, and $\kappa$ are fixed phenomenologically by requiring that the crossover temperature $T_c$ and the pressure $p/T^4$ agree with lattice QCD results at vanishing baryon chemical potential. In this work, the specific values of $T_0, a_0, a_1, a_2, b_3, b_4$, and $\kappa$ are listed in Table I.
\begin{table}
\centering 
\caption{Parameters in the realistic PNJL model~\cite{Bhattacharyya_reparametrizing_2017}.}
\begin{tabular*}{0.48\textwidth}{@{\extracolsep{\fill}} *{4}{c}} 
\thickhline
$m_{u,d}$(MeV) & $m_s$(MeV)& $\Lambda$(MeV) & $G_S\Lambda^2$ \\ 
\hline 
5.5 & 183.468 & 637.720 & 2.914 \\ 
\thickhline
$K\Lambda^5$ & $G_1$(MeV$^{-8}$)& $G_2$(MeV$^{-8}$) \\ 
\hline 
9.496& $2.193\times 10^{-21}$&$-5.890\times 10^{-22}$ \\ 
\thickhline
\end{tabular*}
\end{table}

In Eq. (1), $G_S$ is the scalar coupling constant, and the $K$ term represents the six-fermion Kobayashi-Maskawa-$\prime$t Hooft (KMT) interaction that breaks the $U(1)_A$ axial symmetry~\cite{tHooft_computation_1976}. $\mathcal{L}_1^{8q}$ and $\mathcal{L}_2^{8q}$ are the two eight-quark interaction terms described by~\cite{Bhattacharyya_reparametrizing_2017,Li_the_2019}
\begin{eqnarray}\label{eq4}
\mathcal{L}_1^{8q}=\frac{G_1}{2}\{[\bar{\psi}_i(1+\gamma_5)\psi_j][\bar{\psi}_j(1-\gamma_5)\psi_i]\}^2,
\end{eqnarray}
and
\begin{eqnarray}\label{eq5}
\mathcal{L}_2^{8q}&=&G_2\{[\bar{\psi}_i(1+\gamma_5)\psi_j][\bar{\psi}_j(1-\gamma_5)\psi_k]
\notag\\
&\times&[\bar{\psi}_k(1+\gamma_5)\psi_l][\bar{\psi}_l(1-\gamma_5)\psi_i]\}.
\end{eqnarray}
These two eight-quark terms, governed by coupling constants $G_1$ and $G_2$, resolve the vacuum instability in the (P)NJL model~\cite{Bhattacharyya_reparametrizing_2017}. In addition, these interaction terms can also lower the crossover temperature $T_c$ at vanishing baryon chemical potential, thus bringing them closer to Lattice QCD results and the experimental freeze-out line. The term $\mathcal{L}_{SV}$ is the scalar-vector interaction, whose form can be found in~\cite{Sun_qcd_2021,Yang_properties_2024,Liu_effects_2024} 
\begin{eqnarray}\label{eq6}
\mathcal{L}_{SV}&=&G_{SV}\{\sum_{a=1}^{3}[(\bar{\psi}\lambda^{a}\psi)^{2}+(\bar{\psi}i\gamma_{5}\lambda^{a}\psi)^{2}]\} 
\notag\\
&\times&\{\sum_{a=1}^{3}[(\bar{\psi}\gamma^{\mu}\lambda^{a}\psi)^{2}+(\bar{\psi}\gamma_{5}\gamma^{\mu }\lambda^{a}\psi)^{2}]\}.
\end{eqnarray}

\begin{figure*}[tbh]
\includegraphics[scale=0.44]{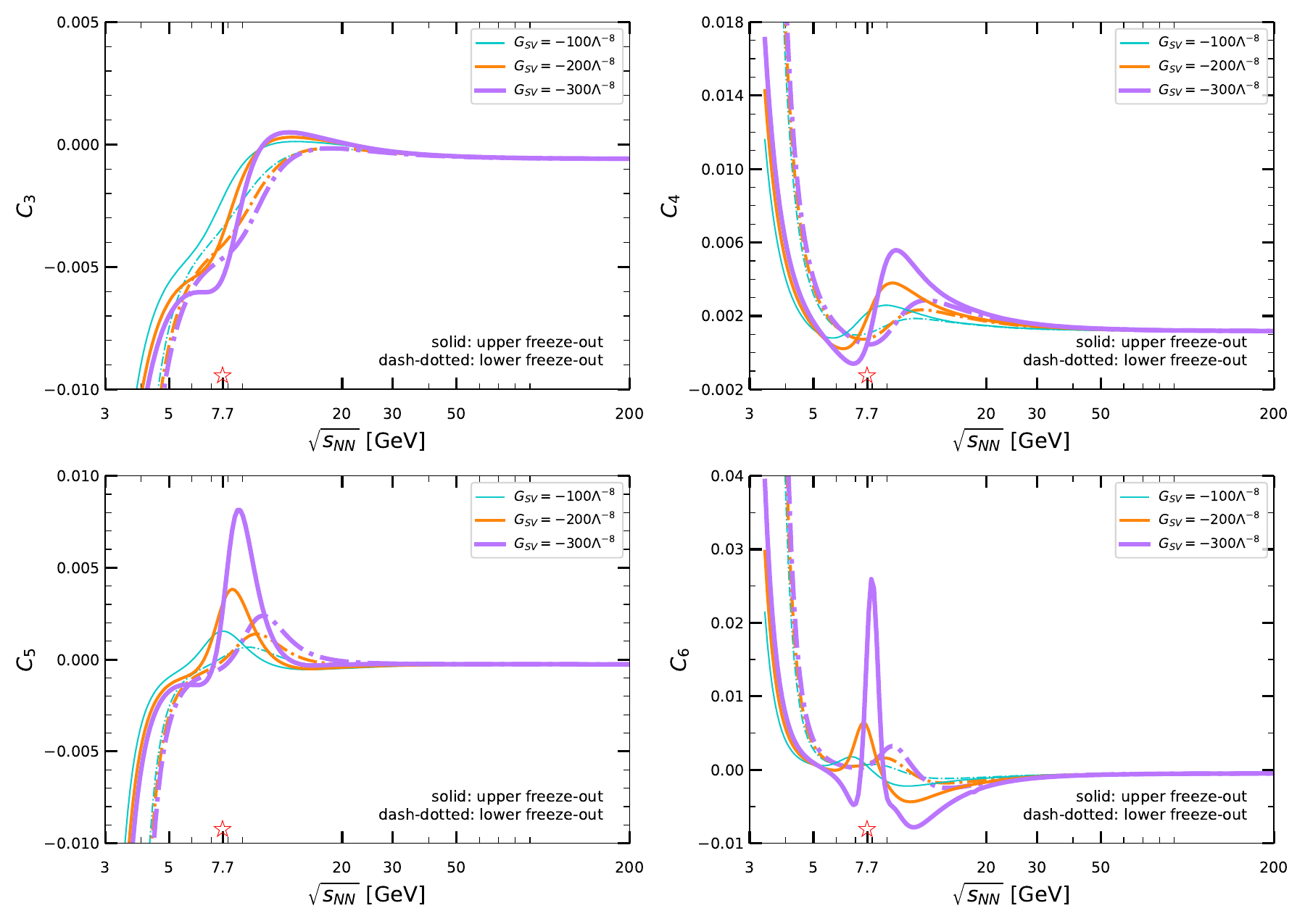}
\caption{Similar to Fig.~\ref{fig2}(b), but for higher-order cumulants of temperature fluctuations $C_3$ through $C_6$.} \label{fig3}
\end{figure*}

Although the scalar-vector coupling interaction has no effect on the QCD vacuum and negligible influence on the phase boundary, the coupling $G_{SV}$ can significantly shift the temperature and baryon chemical potential of CEP~\cite{Sun_qcd_2021}. As demonstrated in Refs.~\cite{Sun_qcd_2021,Sun_spinodal_2022,Yang_properties_2024,Liu_effects_2024}, a negative $G_{SV}$ moves the CEP toward higher $T$ and lower $\mu_B$. Thus, varying $G_{SV}$ enables the chiral phase transition region to overlap with the domain probed in heavy-ion collisions. Accordingly, we take only $G_{SV}$ as a free parameter in this work, while fixing $G_1$, $G_2$, and the thermodynamic-potential parameters (see Table I). The resulting parameter values for the realistic PNJL model are listed in Table II.

In Fig.~\ref{fig1}, we present the phase diagrams in the $\mu_B-T$ plane for different $G_{SV}$ values based on the realistic PNJL model. In the calculations, we adopt the symmetric chemical potential setting $\mu_u=\mu_d=\mu_s=\mu_B/3$. The dashed lines denote the chiral crossover, and the solid diamonds mark the CEPs. Results from functional methods (fRG~\cite{Fu_qcd_2020} and DSE~\cite{Gao_chiral_2021}) and lattice simulations (WB~\cite{Bellwied_the_2015} and HotQCD~\cite{Bazavov_chiral_2019}) are shown for comparison. The chiral crossover temperature is derived from the local maxima of $\partial \langle \bar{u}u \rangle / \partial T$, and the CEP temperature is determined by the intersection of the chiral crossover transition with the spinodal (mechanical) instability region in the $T-n_B$ plane. Fig.~\ref{fig1} demonstrates that the chiral crossover transition is almost insensitive to $G_{SV}$. At vanishing chemical potential, in particular, the critical temperature is consistently $T_c = 156.6$ MeV for all $G_{SV}$ values. Our results are consistent with lattice QCD estimates and comparable to functional-method results. In contrast, a negative $G_{SV}$ significantly raises the temperature of the CEP. At $G_{SV} = -300\Lambda^{-8}$, the CEP is located at $(\mu_B, T)_{\rm CEP} = (447.5, 133)$ MeV. In addition, we also include the experimental chemical freeze-out data (STAR~\cite{Adamczyk_bulk_2017}, Alba et al.~\cite{Alba_freeze-out_2014}, Andronic et al.~\cite{Andronic_decoding_2018}) and freeze-out fitting lines (Lu et al.~\cite{Lu_extracting_2026}, Andronic et al.~\cite{Andronic_the_2010}). The chemical freeze-out conditions are typically determined by fitting particle multiplicities and their ratios within the statistical model. Most of the experimental freeze-out points listed in Fig.~\ref{fig1} lie near the chiral phase transition region predicted by our realistic PNJL model. In particular, the BES-II collision energies of 3–7.7 GeV correspond to a baryon chemical potential range of approximately $760-400$ MeV, which exactly covers the region where the CEP may exist in the model. Therefore, this realistic model, which is consistent with lattice QCD results, is suitable for investigating the characteristics of critical fluctuations.

\textit{Fluctuations of temperature along the freeze-out lines}.
To study temperature fluctuations, Ref.~\cite{Chen_high-order_2025} proposed a systematic theoretical framework by introducing a new thermodynamic function $\Theta  = \Omega + TS$, where $\Omega$ is the thermodynamic potential, and $S$ is the entropy. The differential relation is given by $d\Theta  = TdS -pdV-N_Bd\mu_B$. Consequently, the temperature can be obtained as $T = \partial \Theta /\partial S$, and the $n$-th order temperature fluctuations can be expressed as
\begin{eqnarray}\label{eq7}
\langle (\Delta T)^n \rangle_c = T^{n-1} \frac{\partial^n \Theta }{\partial S^n}=\frac{T^{n-1}}{V^{n-1}} \frac{\partial^n \theta}{\partial s^n},
\end{eqnarray}
with $n\ge 2\ (n\in\mathbb{Z})$, where $\Delta T = T - \langle T \rangle$, $\langle\cdots\rangle$ denotes the ensemble average, $\langle\cdots\rangle_c$ is the cumulant of fluctuations, and the lowercase $\theta$ and $s$ represent the corresponding densities. In Ref.~\cite{Chen_high-order_2025} , the volume is taken as $T^{-3}$, and the dimensionless temperature cumulants $C_n$ are defined as
\begin{eqnarray}\label{eq8}
C_n = \frac{\langle (\Delta T)^n \rangle_c}{T^n} \thicksim T^{3n-4} \frac{\partial^n \theta}{\partial s^n}.
\end{eqnarray}  
Using $\frac{\partial^n \theta}{\partial s^n} = \frac{\partial^{n-1} T}{\partial s^{n-1}}$ and $s = \partial p/\partial T$, $C_n$ can be expressed in terms of the $n$-th derivative of pressure with respect to temperature (see Refs.~\cite{Chen_high-order_2025,Liu_fluctuations_2026}).

\begin{figure*}[tbh]
\includegraphics[scale=0.44]{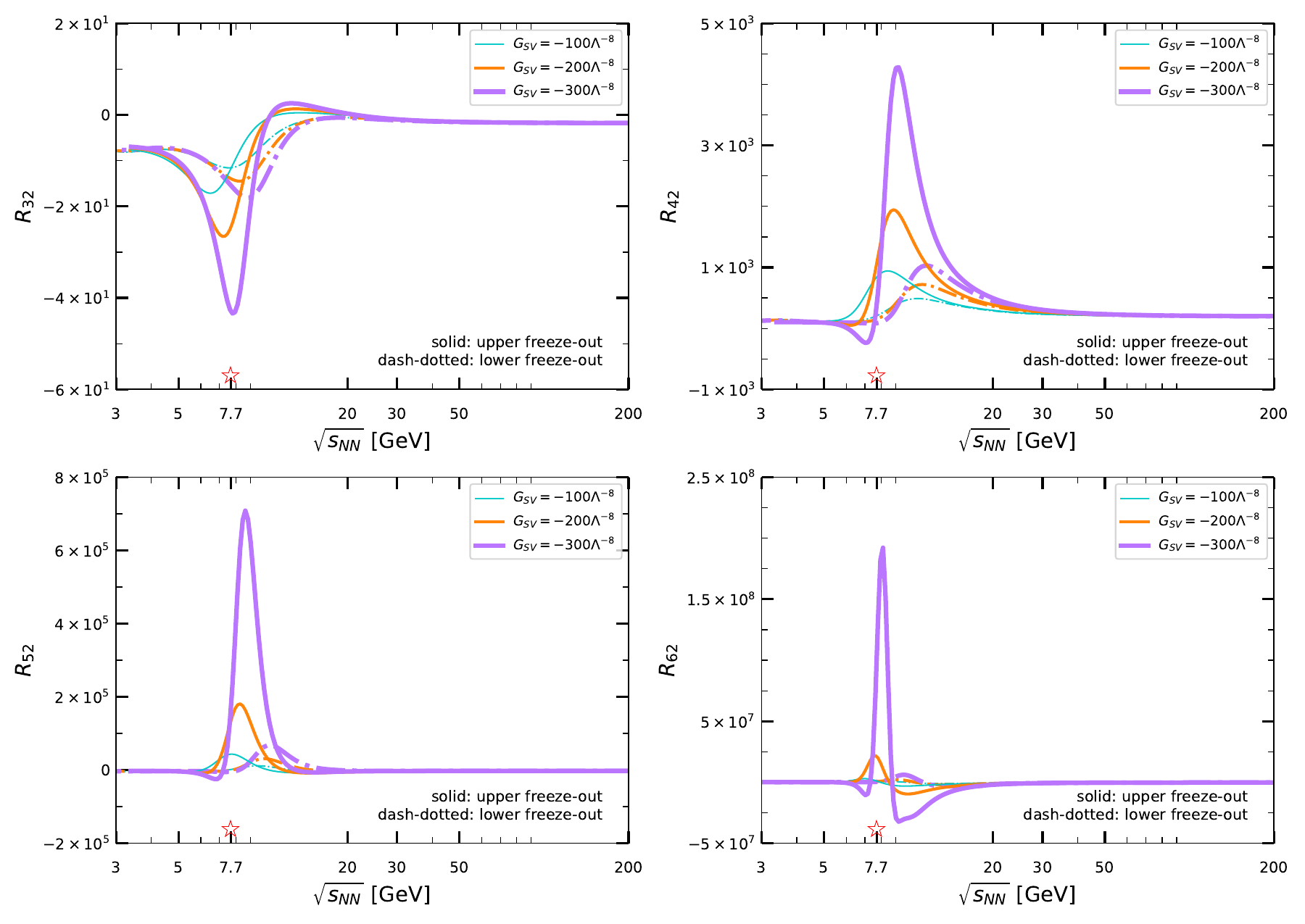}
\caption{Similar to Fig.~\ref{fig2}(b), but for the dimensionless cumulant ratios of temperature fluctuations $C_3/C_2^2$, $C_4/C_2^3$, $C_5/C_2^4$,and $C_6/C_2^5$.} \label{fig4}
\end{figure*}

The variance of temperature fluctuations is characterized by the 2nd-order cumulant $C_2 = T^2 (\partial s/\partial T)^{-1}$, which is precisely the inverse of the dimensionless heat capacity $C_V/T^3$. The two-particle transverse momentum correlation $C_{p_T}$ measured in STAR experiments can be simply taken as $\sqrt{C_2}$. In Fig.~\ref{fig2}(a), we display the contour map of $C_2$ with $G_{SV}=-100\Lambda^{-8}$ in the full $\mu_B$-$T$ phase diagram. As can be seen from the figure, $C_2$ approaches the Stefan-Boltzmann (SB) limit $1/62.5$ at high temperatures or high densities, while it increases rapidly in the low-temperature and low-density region. A pronounced dip structure is clearly observed in the chiral phase transition region, especially near the CEP and the first-order phase transition. The entropy discontinuity at the first-order transition induces a non-monotonic peak in the heat capacity\cite{Bhattacharyya_finite_2020}, which in turn suppresses $C_2$ sharply toward zero, forming the dip structure. The speed of sound $c_s^2$ exhibits a similar behavior, dropping to near zero around the CEP and the first-order transition as well~\cite{Liu_speed_2024}. 

The STAR Collaboration has reported the observation of a dip structure in central (0–5\%) collisions at beam energies of 3–7.7 GeV using the high-statistics BES-II data. Accordingly, in Fig.~\ref{fig2}(b), we plot $C_2$ along the experimental chemical freeze-out lines. Experimentally, the freeze-out conditions are typically determined by fitting particle yields and their ratios within the statistical model~\cite{Andronic_decoding_2018}. As shown in Fig.~\ref{fig1}, the currently available freeze-out data and their fitted lines (e.g. Andronic et al.~\cite{Andronic_the_2010}) are mainly based on BES-I and SPS data, lacking the high-precision BES-II statistics. Recently, Ref.~\cite{Lu_extracting_2026} reported a self-consistent \textit{functional} extraction of freeze-out conditions from STAR data, covering BES-I (200-27 GeV), BES-II (27-7.7 GeV), and SPS experiments, thereby providing a more reliable reference in the intermediate-\(\mu_B\) region. Therefore, we adopt this set of freeze-out conditions in the present work. The purple band in Fig.~\ref{fig1} represents the predicted freeze-out conditions, whose upper and lower boundaries are parameterized as~\cite{Cleymans_comparison_2006,Lu_extracting_2026} 
\begin{eqnarray}\label{eq9} 
T_f = T_0 \left[ 1 - \kappa_2^f \left( \frac{\mu_{B,f}}{T_0} \right)^2 - \kappa_4^f \left(\frac{\mu_{B,f}}{T_0} \right)^4 + \cdots \right],
\end{eqnarray} 
where $T_0$ is the freeze-out temperature at zero chemical potential, and the freeze-out $\mu_{B,f}$ is related to the collision energy via $\mu_{B,f} = a/(1 + b\sqrt{s_{NN}})$. The fit parameters $T_0$, $a$, $b$ and curvature coefficients $\kappa_2^f$, $\kappa_4^f$ are listed in Table III~\cite{Lu_extracting_2026}.

\begin{table}
\centering 
\caption{Parameters for the upper and lower freeze-out bounds~\cite{Lu_extracting_2026}.}
\begin{tabular*}{0.48\textwidth}{@{\extracolsep{\fill}} *{6}{c}} 
\thickhline
& $a$(MeV)& $b$(GeV$^{-1}$) & $T_0$(MeV) & $\kappa_2^f$ & $\kappa_4^f$\\ 
\hline 
upper & 1027.0 & 0.2143 & 157 & 0.0153 & $0.73\times 10^{-3}$\\ 
\hline 
lower & 913.9 & 0.1977 & 157 & 0.0153 & $1.47\times 10^{-3}$ \\ 
\thickhline
\end{tabular*}
\end{table}

In Fig.~\ref{fig2}(b), we plot $C_2$ along the upper and lower freeze-out lines for various $G_{SV}$. All curves exhibit a non-monotonic dip structure when the freeze-out line passes near the chiral phase boundary, and the dip is more pronounced for the upper freeze-out, which is closer to the boundary. A negative $G_{SV}$ enhances the CEP temperature, moving the freeze-out line closer to the CEP and deepening the dip minimum. In general, irrespective of parameters or model details, a freeze-out line nearer the CEP yields a stronger dip. In our realistic PNJL model, the dip occurs around 7.7 GeV and follows the trend observed by STAR experiment. Therefore, our results suggest that the observed dip in $C_{p_T}$ could be associated with the CEP. However, it should be noted that our theoretical model is based on the equilibrium assumption and performed in the mean-field approximation，thus ignoring quantum fluctuations. At low-energy collisions, volume fluctuations become very significant due to the lower multiplicity, yet they are not included in Eq.~\ref{eq8}. As a result, while our results reproduce the non-monotonic dip trend, quantitative deviations remain. To further determine whether the mean $\langle p_T \rangle$ fluctuations are related to the CEP, the STAR collaboration is advancing measurements of higher-order cumulants~\cite{Gao_Collision_2025}. In Fig.~\ref{fig3}, we present our predictions for $C_3$ through $C_6$ along different freeze-out lines. The $C_3$ results are negative overall and do not show significant non-monotonic behavior, while $C_4$ exhibits a clear peak-dip non-monotonic structure that could be detected experimentally. Although $C_5$ and $C_6$ also display notable non-monotonic variations, precise experimental measurements of these higher-order cumulants may face greater challenges.

Cumulant ratios, e.g., $C_4/C_2$ for net-proton numbers, are a common strategy to eliminate initial volume fluctuations~\cite{Aboona_precision_2025}. For the temperature fluctuations, the dimensionless cumulant ratios $R_{n2}$ are defined as
\begin{eqnarray}\label{eq10} 
R_{32}=\frac{C_3}{C_2^2}, \ R_{42}=\frac{C_4}{C_2^3}, \ R_{52}=\frac{C_5}{C_2^4}, \ R_{62}=\frac{C_6}{C_2^5}.
\end{eqnarray} 
The relevant cumulant ratios in Fig.~\ref{fig4} reveal significant non-monotonic behavior, including a pronounced main peak (or dip). This feature arises primarily from the denominator $(C_2)^{n-1}$, which drops rapidly to a minimum or even approaches zero near the CEP. The absence of a dip structure in $R_{52}$ is due to the non-monotonic variation of $C_5$ near the CEP, which produces a positive peak. When the freeze-out line is closer to the CEP, for example when the CEP for $G_{SV}=-300\Lambda^{-8}$ is the nearest to the upper freeze-out, the absolute values of all main peaks (dips) systematically increase. The cumulant ratios of temperature fluctuations accordingly show a strong correlation with the CEP. As a result, we argue that cumulant ratios like $C_3/C_2^2$ or $C_4/C_2^3$ are more indicative of the relation between $C_{p_T}$ and critical thermodynamic fluctuations.

\textit{Conclusions and outlook.} 
In this work, we extend and reparametrize the PNJL model by introducing two eight-quark interactions to reproduce lattice QCD data at zero baryon chemical potential, and a vector-scalar coupled interaction to adjust the CEP position so that it is covered by the BES-II high-statistics data at collision energies of 3–7.7 GeV. To understand the non-monotonic energy dependence of the two-particle transverse momentum correlation $C_{p_T}$ reported by the STAR Collaboration, we employ this realistic model to investigate the phase-diagram behavior of the second-order temperature fluctuation cumulant $C_2$ and its evolution along the freeze-out lines. Our results indicate a distinct dip in $C_2$ in the chiral phase transition region, especially near the CEP and the first-order phase boundary. When the experimentally extracted chemical freeze-out line crosses the chiral transition region of the phase diagram, the dip becomes more pronounced as the line approaches the CEP. In our realistic PNJL model, the dip occurs around 7.7 GeV and follows the trend observed by the STAR experiment. This suggests that the observed dip in $C_{p_T}$ could be associated with the CEP. We also present the non-monotonic collision-energy dependence of higher-order cumulants. Although multiple parameter sets are introduced to bring the freeze-out line closer to the CEP, they do not affect the qualitative fluctuation features. However, our PNJL framework is based on equilibrium thermodynamics and the mean-field approximation, and neglects initial volume fluctuations in $C_2$. Consequently, while the trend and minimum position agree with experiment, the numerical magnitudes deviate substantially. To eliminate initial volume fluctuations, we also examine the dimensionless cumulant ratios of temperature fluctuations, which exhibit prominent peaks (or dips) in their non-monotonic variations. We argue that cumulant ratios like $C_3/C_2^2$ or $C_4/C_2^3$ are more indicative of the relation between $C_{p_T}$ and critical thermodynamic fluctuations. Although doubts persist regarding the critical origin of the observed non-monotonicity~\cite{Reichert_non-monotonicity_2026}, STAR is progressing toward higher-order event-mean $\langle p_T \rangle$ fluctuation measurements~\cite{Gao_Collision_2025}, and theoretical confirmation could be further pursued through hydrodynamic~\cite{Du_kinematic_2026} or transport simulations~\cite{Cao_specific_2022,Zhang_energy_2025} that incorporate critical fluctuations~\cite{Zhou_elliptic_2021}.

\textit{Acknowledgments.}
We gratefully acknowledge Chunjian Zhang and Wen-Hao Zhou for insightful discussions. We thank Yi-Hang Xing for discussions on the calculation of intensity fluctuations. This work was supported by the National Natural Science Foundation of China (Grant Nos. 12205158 and 12575134), the Shandong Provincial Natural Science Foundation, China (Grant Nos. ZR2021QA037, ZR2022JQ04, and ZR2025QC1487), and the Qingdao Natural Science Foundation (Grant No. 25-1-1-4-zyyd-jch).

\end{document}